\documentclass[smallextended]{svjour3}       % onecolumn (second format)
\smartqed  % flush right qed marks, e.g. at end of proof
\usepackage{graphicx}
\usepackage{graphicx}
\usepackage{subcaption}
\usepackage{mwe}
\usepackage{colortbl}
\usepackage{hyperref}
\usepackage[utf8]{inputenc}

\begin{document}

\title{Turing Award elites revisited: patterns of productivity, collaboration, authorship and impact %\thanks{Grants or other notes
}


\author{Yinyu Jin\textsuperscript{1}         \and
        Sha Yuan\textsuperscript{1$*$} \and
        Zhou Shao\textsuperscript{2, 4}  \and
        Wendy Hall\textsuperscript{3}  \and
        Jie Tang\textsuperscript{4}%etc.
}


\institute{
$^*$ Sha Yuan is the corresponding author.  
              \email{yuansha@baai.ac.cn}   \\
    \textsuperscript{1} Beijing Academy of Artificial Intelligence, Beijing, China \\
    \textsuperscript{2}  Nanjing University of Science and Technology, Nanjing, China \\
    \textsuperscript{3} University of Southampton, Southampton, U.K. \\
    \textsuperscript{4} Department of Computer Science and Technology, Tsinghua University, Beijing, China
}

\date{Received: date / Accepted: date}

\maketitle

\begin{abstract}
The Turing Award is recognized as the most influential and prestigious award in the field of computer science(CS). With the rise of the science of science (SciSci), a large amount of bibliographic data has been analyzed in an attempt to understand the hidden mechanism of scientific evolution. These include the analysis of the Nobel Prize, including physics, chemistry, medicine, etc. In this article, we extract and analyze the data of 72 Turing Award laureates from the complete bibliographic data, fill the gap in the lack of Turing Award analysis, and discover the development characteristics of computer science as an independent discipline. First, we show most Turing Award laureates have long-term and high-quality educational backgrounds, and more than 61\% of them have a degree in mathematics, which indicates that mathematics has played a significant role in the development of computer science. Secondly, the data shows that not all scholars have high productivity and high h-index; that is, the number of publications and h-index is not the leading indicator for evaluating the Turing Award. Third, the average age of awardees has increased from 40 to around 70 in recent years. This may be because new breakthroughs take longer, and some new technologies need time to prove their influence. Besides, we have also found that in the past ten years, international collaboration has experienced explosive growth, showing a new paradigm in the form of collaboration. It is also worth noting that in recent years, the emergence of female winners has also been eye-catching. Finally, by analyzing the personal publication records, we find that many people are more likely to publish high-impact articles during their high-yield periods.

\end{abstract}

\section{Introduction}
\label{intro}

Science is the ladder to the progress of human civilization. From the perspective of scientometrics, science can be described as a complex, self-organizing, and evolving network of scholars, projects, papers, and ideas \cite{fortunato2018science}. Science of science (SciSci), as an emerging field, is based on numerous digital bibliographic data. It tries to dig out the underlying mechanism of the genesis of scientific discovery, reveal its structure and influencing factors, and quantitatively understand the evolution of science. In this way, it allows us to develop predictive models to capture the emergence of scientific discoveries, improve individual scientist's success, and enhance the outlook of science as a whole \cite{fortunato2018science}. The booming of SciSci benefits from two aspects. One is the digitization of scholarly information separated from traditional paper-based media and more and more accessible digital data sources from the Internet nowadays (Aminer \cite{tang2008arnetminer}, Microsoft Academic \cite{sinha2015overview}, Web of Science (WoS), Scopus, Google scholar \cite{van2014google}, and others), some of which are publicly available. This has become the essential fundamental data sets that we can utilize to conduct research. The second is the development of a wide range of quantitative methods, such as descriptive statistics, data visualization, network science methods, machine learning algorithms, mathematical analysis, and computer simulation, including agent-based modeling \cite{fortunato2018science}.

With the popularity of bibliographic data, more and more publicly accessible data can be used to analyze the scholars' work. It enables a higher insight to see the evolution of distinct work and even attempts to predict critical events, such as who, when, and where there may be potentially significant discoveries in the future, assist the decision-makers of crucial decisions. This is the reason for the rise and vigorous development of the SciSci in recent years. At present, some studies \cite{stephan1993age}\cite{jones2011age}\cite{wagner2015nobel}\cite{li2020scientific} have been conducted on the Nobel Prize in the hard sciences, covering Physics, Chemistry, Physiology or Medicine. However, the Nobel Prize does not involve computer science; instead, the Turing Award is regarded as the "Nobel Prize of Computing". The A.M. Turing Award was established by the American Computer Association in 1966 to reward individuals who have made outstanding contributions of lasting importance to the computer industry. It was named in honor of the pioneer of Alan Mathison Turing (1912-1954), a British mathematician and computer scientist. The Turing Award has exceptionally high requirements for the winners, and the award procedure is stringent. Generally, only one computer scientist is awarded each year. Only a minimal number of years have more than two scientists who contribute in the same direction. To date, there is no comprehensive and in-depth analysis of computer science. Therefore, we conduct this research for the Turing Award winners, representing the top elites in the computer field.

\section{Data}

\begin{figure}[ht]
  \includegraphics[width=1\textwidth]{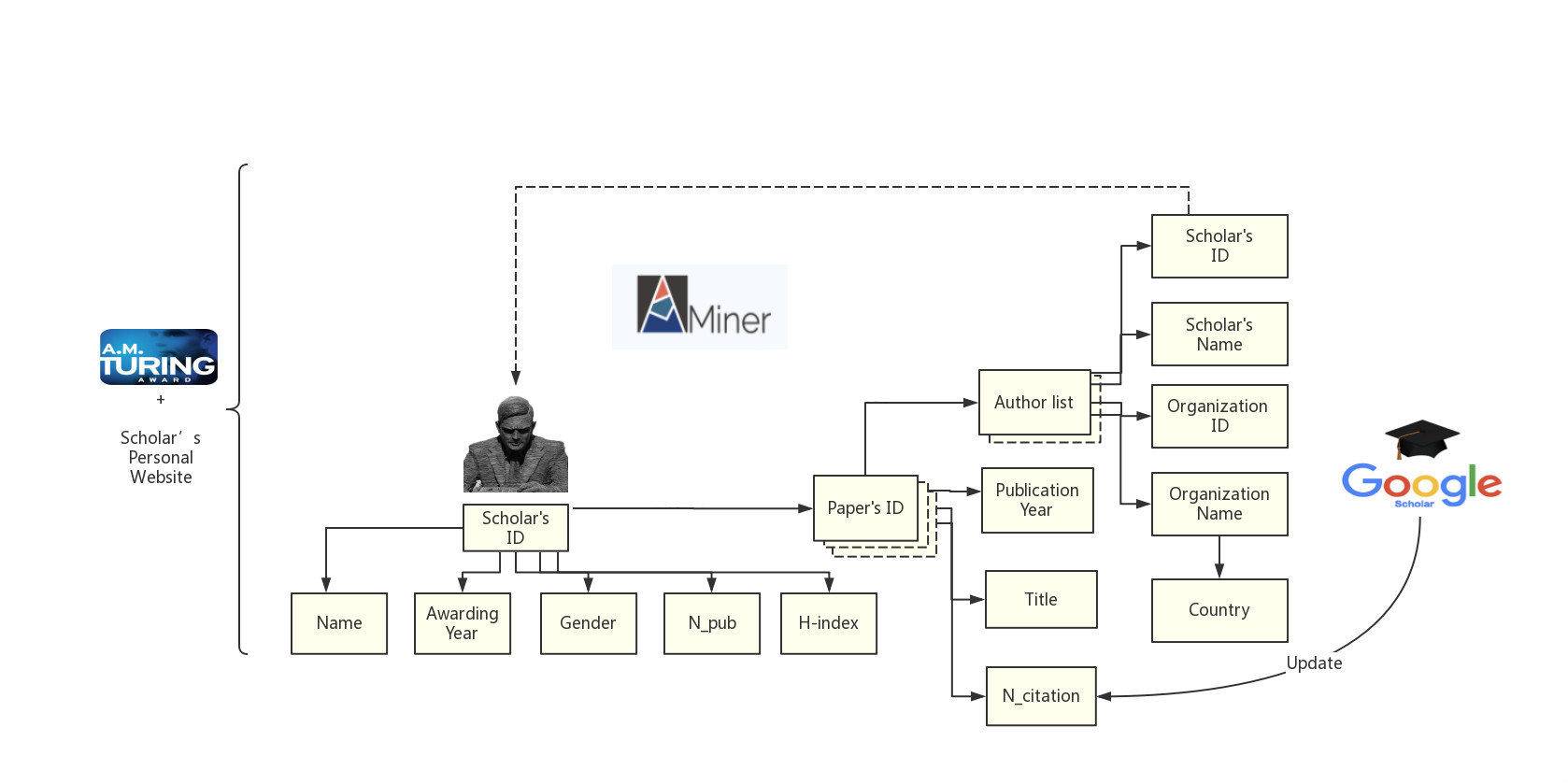}
\caption{We select Aminer data set as our fundamental database. Then the data set is reconstructed manually according to the Turing official website, scholar's personal website, and Google Scholar.}
\label{fig:data}       % Give a unique label
\end{figure}

Here, we compared some large publicly available datasets: Scopus, Web of Science (WoS, formerly known as Web of Knowledge), Aminer dataset, and MAG dataset. First, Scopus covers a comprehensive range of citations, including the Institute for Scientific Information's (ISI) citation database and trade journals. Since we focus more on the academic aspects, such as the Science Citation Index (SCI), the Scopus database is beyond the scope and not used directly. Second, WoS was originally produced by the ISI, but citations are more focused on 'high-influence' rather than comprehensive, for example, less coverage of publications outside the U.S., non-English language publications, and interdisciplinary field publications. That is, "should not be used alone to find citations to authors or titles"\cite{yang2006citation}. In the end, we choose the latest Aminer dataset because it is the most recent update (2020-04-02) and contains more field information than MAG. Also, we verified that most of Aminer's data (such as the publication list) are more comprehensive than WoS. AMiner provides free literature data. The system has been running on the Internet since 2006. It can automatically extract scholars' data from the Internet and integrate the published data in online databases such as DBLP, ACM Digital Library, and CiteSeer.

We extract a list of 72 Turing laureates and their publication list from the Aminer database. Then, the data set is reconstructed according to Google Scholar, Turing official website\footnote{\url{https://amturing.acm.org/}} and scholars' personal websites ( Fig. \ref{fig:data}). For example, some "null" fields are supplemented, and some deviation information is manually corrected to ensure more accurate analysis results. It is essential to obtain the most accurate and comprehensive data set as possible because our research topic is limited to the elites of computer science, and the research results are susceptible to the data set. As a study of SciSci, we conducted a qualitative and quantitative analysis on it. Our complete data set can be downloaded from the website\footnote{\url{http://open.baai.ac.cn/data-set-detail/MTI2NDk=/NjY=/true}}.

As of the publication of this paper (2020), ACM(Association for Computing Machinery)\footnote{\url{https://www.acm.org/}} has conferred 72 laureates from all over the world who have made outstanding contributions to the computer field in 55 years from 1966 to 2020. The number of total publications of these laureates is 13,793 in our data set. 

\section{Result}

\subsection{Overview}

\subsection{Universities, Education, and Majors}

\begin{figure}[ht]
  \includegraphics[width=1\textwidth]{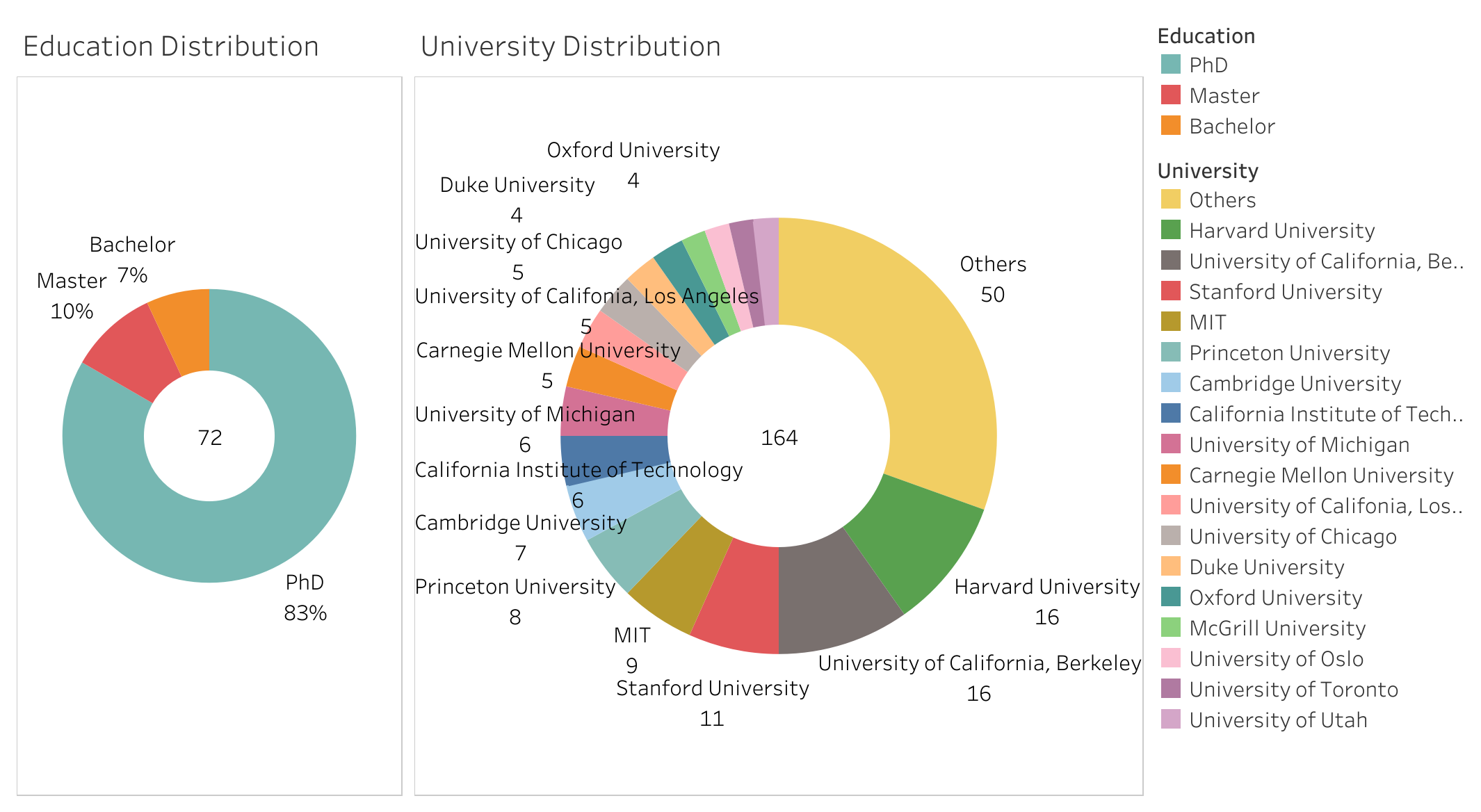}
\caption{These two pie charts capture the proportion of the different levels of education of the total laureates and the proportion of the universities where they obtained their academic degrees. The number of universities is calculated for each laureate at each education period. For example, if one laureate gets his bachelor's degree in institution A, and a master's degree and a PhD in institution B, then A is counted once, and B is counted twice.}
\label{fig:edu}       % Give a unique label
\end{figure}

In Fig. \ref{fig:edu}, we counted the universities where the awardees studied and obtained degrees, including undergraduates, masters, and doctoral degrees. If a scientist has two degrees in a university, such as a bachelor's degree and a doctorate, then the university is counted as two degrees. Our results show that Harvard University and the University of California, Berkeley (UCB) contributed the most degrees with 16 each, followed by Stanford University with 11, and then Massachusetts Institute of Technology (MIT) (9), Princeton University (8), Cambridge University (7), and California Institute of Technology (6), University of Michigan (6), Carnegie Mellon University (CMU) (5), University of California, Los Angeles (UCLA) (5), University of Chicago (5), etc. Obviously, most of these high-frequency universities are ranked in the top 30 of the major university rankings\footnote{These 11 universities are ranked in the top 30 of the three major world university rankings (QS2020, THE2020, ARWU2019), except for one in the ARWU2019 ranking and two in the QS2020 ranking; they all ranked in the top 100 in these three major world university rankings.}. This shows that most of these awardees are the ones with outstanding learning ability because admission to these universities requires great competitiveness.

Among the 72 laureates, 83\% received a PhD (60), 10\% received a master's degree (7), and 7\% obtained a bachelor's degree (5). It demonstrates that almost all of them have accumulated a large amount of long-term academic knowledge, most of which have obtained doctoral degrees, and those with master's and undergraduate degrees have also devoted their whole lives to their research fields. This further shows that these elites in CS are not only outstanding students with strong learning abilities but also invested an enormous amount of time and energy in accumulating and exploring their own subjects. Therefore, the Turing Award is not obtained by luck but by continuous effort and dedication driven by an intense thirst for knowledge and exploration. 

Interestingly, by analyzing the majors during their education, we find that 44 of 72 laureates have degrees in mathematics, then followed by electrical engineering, computer science, physics, and mechanics. This indicates that mathematics plays a vital role in the foundation and essential parts of CS. Besides, there are also majors like nuclear engineering (Patrick M. Hanrahan, 3D computer graphics), biophysics (Patrick M. Hanrahan, 3D computer graphics) and political science (Herbert A. Simon, a basic contributor to artificial intelligence, the psychology of human cognition, and list processing), civil engineering (Dabbala Raj Reddy, design and construction of large scale artificial intelligence systems), psychology (Geoffrey E. Hinton, deep neural networks) and astronomy (Peter Naur, programming language design to compiler design).  

\subsubsection{Country Distribution and Glittering of Women Impact}

\begin{figure}[ht]
  \includegraphics[width=1\textwidth]{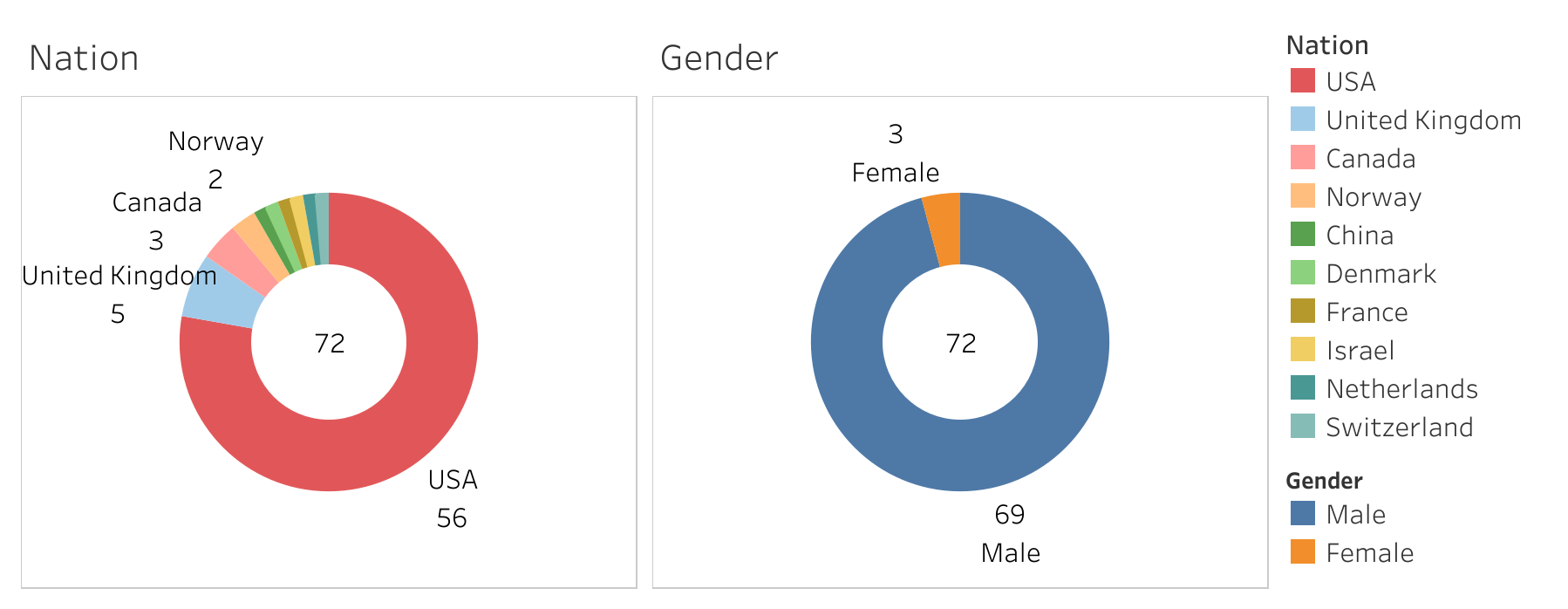}
\caption{These two pie charts capture the proportion of the nationality and gender of the laureates.}
\label{fig:2}       % Give a unique label
\end{figure}

In this section, the origin countries are assigned based on the nationality of the Turing Award winners at the data retrieval time. In Fig. \ref{fig:2}, like the country that has nurtured the most Nobel Prize winners \cite{tyutyunnik2013scientometric}, the United States is also the main contributor to Turing Award, with 56 winners, accounting for 77.8\%. The UK ranked second with five laureates, followed by Canada and Norway with 3 and 2 respectively. Some other European countries (Denmark, France, the Netherlands, and Switzerland) and two Asian countries (China and Israel) also each have a winner. Noticeably, in the proportion of the United States, few laureates are immigrants or joint citizenship. For example, Dabbala Raj Reddy, born in India, completed his PhD in speech recognition under McCarthy's\footnote{John McCarthy: Turing Award laureate in 1971 for considerable contributions to the foundation of artificial intelligence.} guidance at Stanford University; Silvio Micali, born in Italy, earned his PhD at the University of California, Berkeley and worked at MIT; with both Israeli parents, Shafi Goldwasser is born and educated in the United States; Judea Pearl, also with both Israeli parents, started his graduate study and research work in the United States.
Thus, it also implies that the U.S. is still the main birthplace of computer talents.

In the field of computer science, the underrepresentation of women is a concerning phenomenon. According to the literature, in 1990, only 13\% of PhDs in CS were women, and only 2.7\% of CS tenured professors were women \cite{spertus1991there}. Statistics from 1980 to 1994 also show an incredible shrinking pipeline for women in CS \cite{camp2002incredible}. This gender difference is the result of social and structural factors. Social and cultural biases incorporate both the women's internal view of themselves (self-expectations) and the external view (stereotypes) generally held by society\cite{ahuja2002women}. In terms of structural factors, the scarcity of women role models and mentors and the low proportion of women in the top ranks are examples.

It is also reflected in the Turing Award. There are only three female laureates, accounting for 4\% of the total of 72. However, despite the 40-year blank since the Turing Awards began, the emergence of three female researchers in 2006, 2008, and 2012 attracted public attention, which means that women are getting involved in computer science and making a significant impact on the global stage. Frances Allen was inspired by John Backus'\footnote{John Backus: Turing Award laureate in 1977 for contributions to the design of high-level programming systems, notably on FORTRAN.} work in the compiler, also collaborated with John Cocke\footnote{John Cocke: Turing Award laureate in 1987 for contributions in the design and theory of compilers.}, and won the Turing Award in 2006 for the contribution to the theory and practice of optimizing compiler techniques in IBM at the age of 74. She also worked to encourage other women to participate in computer-related fields. Barbara Liskov was awarded for her work in theoretical and practical foundations of programming language and system design in her 69. At the same time, she is one of the first women in the United States to receive a PhD in computer science. Shafi Goldwasser, joint American/Israeli citizenship, along with Silvio Micali were awarded the prize in 2012 for their work on cryptography, computational complexity, computational number theory, and probabilistic algorithms at the age of 53. Some efforts have been carried out, aiming to increase women's participation in CS \cite{roberts2002encouraging}. However, on a global scale, the aforementioned social and structural problems remain severe. To eventually realize the revolution, computer scientists and educators need to make dramatic changes and take concrete actions in the direction \cite{camp2002incredible}.

\subsubsection{Publications and h-index}

\begin{figure}[ht]
  \includegraphics[width=1.2\textwidth]{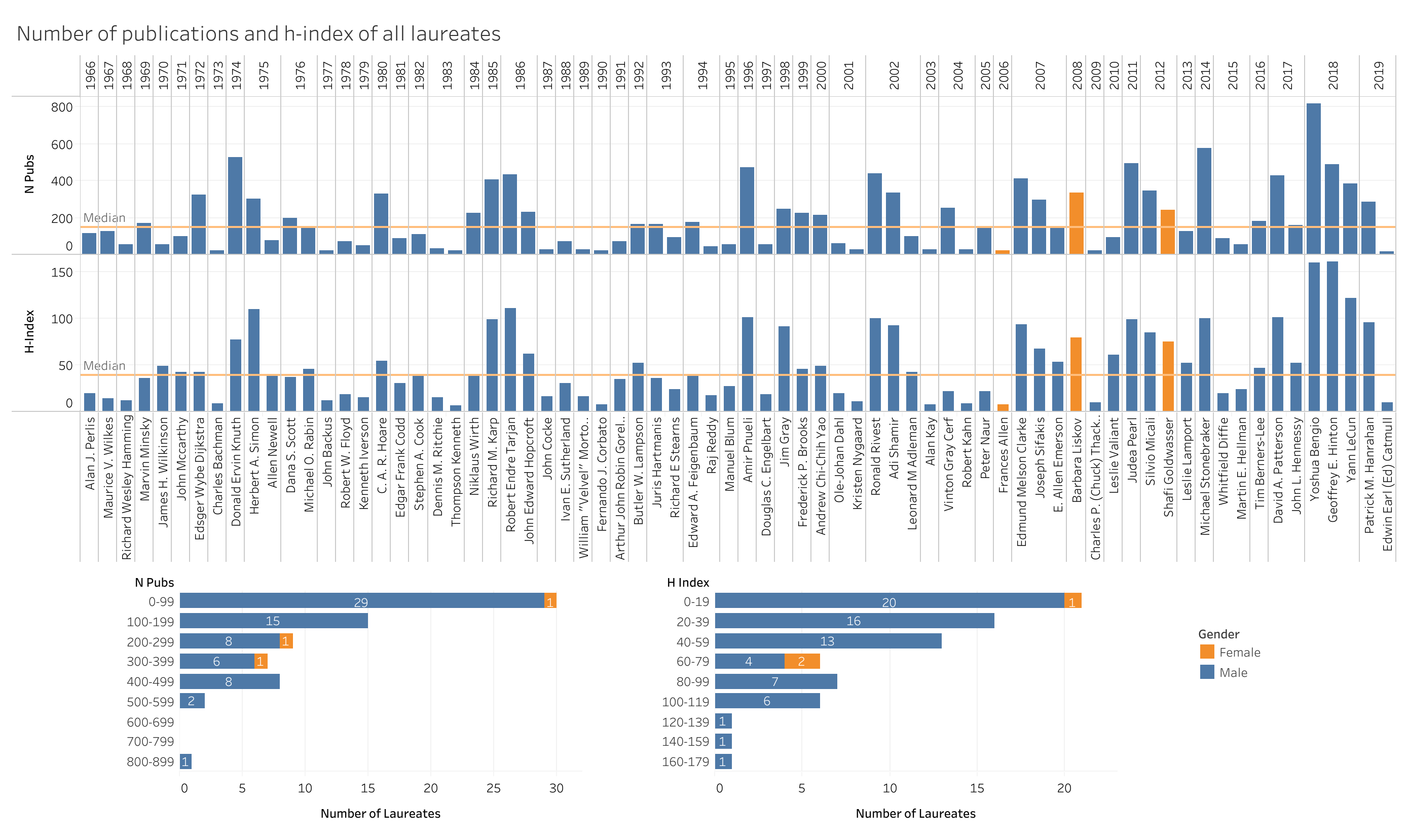}
\caption{The total number of publications and the h-index of each laureate are sorted by the award year. The female laureates are colored in orange, while the males are in blue. The orange horizontal line is the median reference line of the two indicators. The two charts below calculate the number of winners in each interval.}
\label{fig:1}       % Give a unique label
\end{figure}

According to our reassembled data set, we plot histograms of the total number of publications and h-index for all the laureates in the order of the awarding years. H-index \cite{je2005index} is the most widespread measurement of achievement indicators for researchers. In Fig. \ref{fig:1}, the median number of papers published by each researcher is 147.5, and the median h-index is 39. We found that the distribution of these two indicators is similar, demonstrating that the more articles published, the higher the h-index. According to the definition of h-index, which captures both quantity (number of publications) and quality (citations for each paper), we can infer that the number of high-impact papers published by them is sufficient to compare with the number of total publications.

In addition, the productivity of these researchers varies greatly, ranging from 16 to more than 800. The standard deviation and coefficient of variation of the number of publications were 167.9 and 88.25\%, respectively. We see that almost half of them published no more than 100 papers. The other half published more than 100 papers. We also find an extreme number of publications of 816, Yoshua Bengio, reaching the h-index of 159. It is worth noting that the three 2018 laureates who jointly published "Deep Learning" in 2015 are among the top on both charts and are far ahead of other researchers in terms of h-index, all exceeding 120. In fact, these indexes are even higher than science giants like Einstein, Darwin, and Feynman, who have impressive h-index of 96, 63, and 53, respectively, in the 2012 publication \cite{acuna2012predicting}. This is because the exponentially increasing number of publications \cite{fortunato2018science}, as well as the corresponding citation trend from 1900 to 2018, also positively influenced the h-index upper limit. 

Both of these statistics follow the Pareto-like distribution. 40\% of the researchers published 80\% of the paper, and the majority of the remaining people accounted for 20\%, similar to the Pareto principle. This shows that Pareto's law still applies to the elite world. There is an exponentially growing trend of the annual number of articles from 1900 to 2018 for all disciplines \cite{fortunato2018science}. However, in our computer science elites world, this trend is not apparent. 

Hence, we separately analyzed the bottom 10 least publications winners \footnote{Ten laureates with the least number of publications: Edwin E. Catmull, Fernando J Corbato, Frances E Allen, Kenneth L Thompson, Charles P. Thacker, John Backus,  Charles W Bachman, John Cocke, William M Kahan, Alan Kay} and the top 10 most publications winners \footnote{Ten laureates with the highest number of publications: Yoshua Bengio, Michael Stonebreaker, Donald E Knuth, Judea Pearl, Geoffery E Hinton, Amir Pnueli, Ronald, Robert Tarjan, David Patterson, Edmund M Clarke}. First, for the ten people with fewer publications, half of them have MS/BS degrees, accounting for 5/12 of the total MS/BS degrees. Eight people (80\%) are in the industry, such as IBM, Bell Labs, and Microsoft. They are dedicated to the fields such as programming languages, computer architectures, and operating systems. On the other hand, for the top 10 winners of most publications, all of them have a PhD and are committed to academic research in areas such as artificial intelligence, data structures, and theories. This shows that engineering and theory are equally important in the computer field, and not all the works are represented in the form of publications. 

\subsection{Awarding Age and the Number of Years Experienced}

\begin{figure}[ht]
  \includegraphics[width=1.2\textwidth]{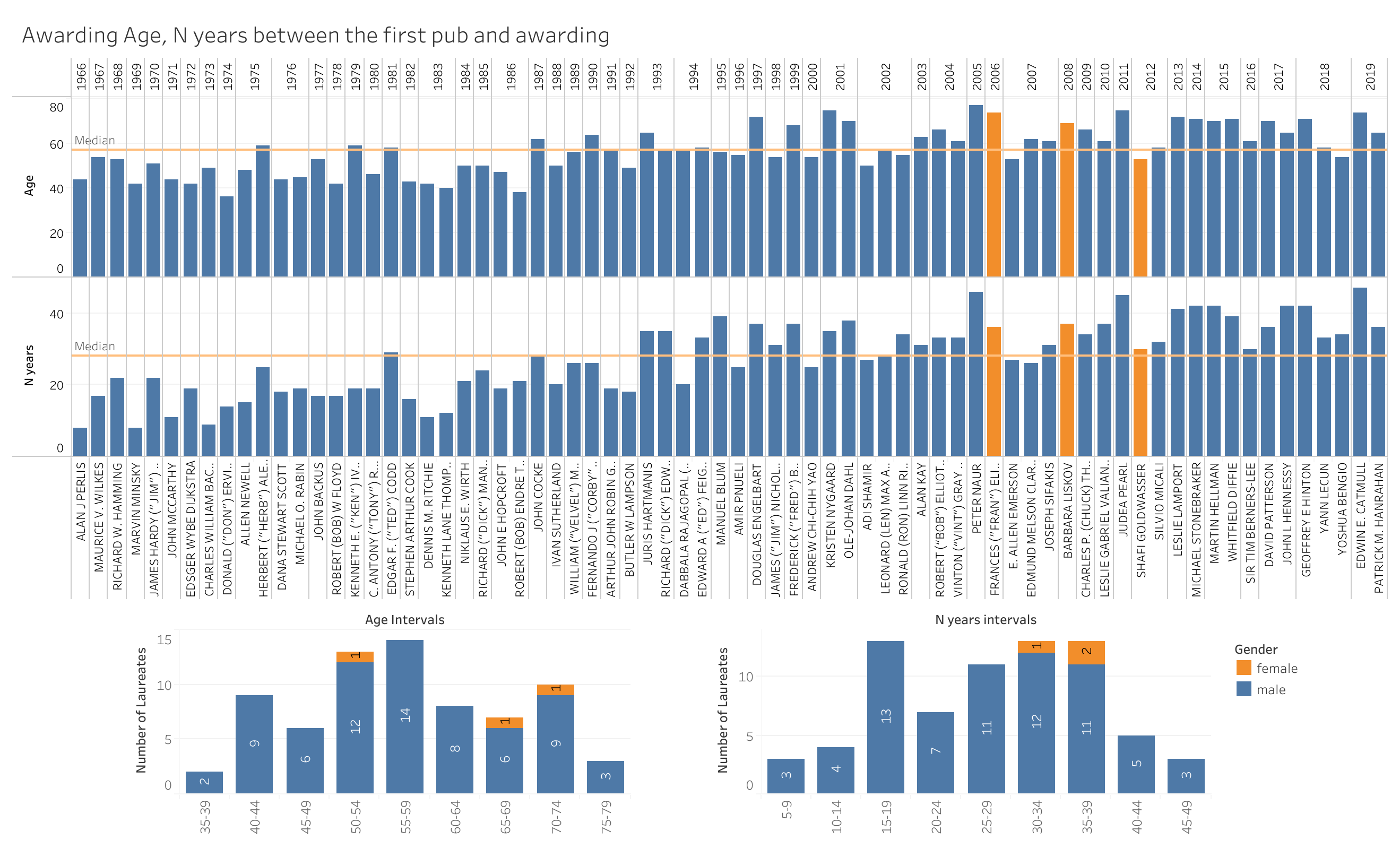}
\caption{The diagram captures the age of awarding and the number of years from the first publication to the awarding year in the sequence of the awarding year. The orange line is the median reference line of the axis. Female laureates are colored in orange, while males are in blue. And the diagram below counts the number of laureates in each interval.}
\label{fig:3}       % Give a unique label
\end{figure}

Judging from the age of the scientists at the year of the award, the youngest winner was Donald E. Knuth in 1974, at the age of 36; the oldest winner was Peter Naur in 2005, was 77 years old (see Fig. \ref{fig:3}). The median age of scientists at the time of the award is 57, and most are mainly distributed between 40-74 years old, especially in their 50s. The histogram has an evidently increasing trend, where before 1987, most of the laureates are about 40 years old, and from 1987 to 2002 the mean age is near the total median line of 57, after 2002 most of the laureate's age exceeded the median reference line reached about 70 years old. To clarity the age-creativity relationship, we analyzed the laureates of the most recent decade (2010-2019) and found that their pioneering work was not immediately reflected in the award, but many years, even two decades, later. This may be because significant research impact takes time to verify, and maybe the Turing Award also focuses on one's lifelong commitment and contribution, rather than just a 'flash in the pan'.

Second, we plot the number of years from the author's first publication to the awarding year. According to statistics, the median number of 28 years is needed from their first publications. At the same time, this distribution also has a clear upward trend. In the early years of the Turing Awards, the laureates received the award in less than 20 years. By contrast, this number has been increasing since 1993, reaching about 40 years. There are probably various factors behind this phenomenon. However, an expansion of fundamental knowledge in a field may increase training requirements, making it more difficult to contribute at a young age \cite{jones2009burden} \cite{jones2010age}. This also indicates that the Turing Award is more likely to grant researchers who make a profound and lasting contribution in this research field.

\subsection{Research Subject}

\begin{figure}[ht]
  \includegraphics[width=1.2\textwidth]{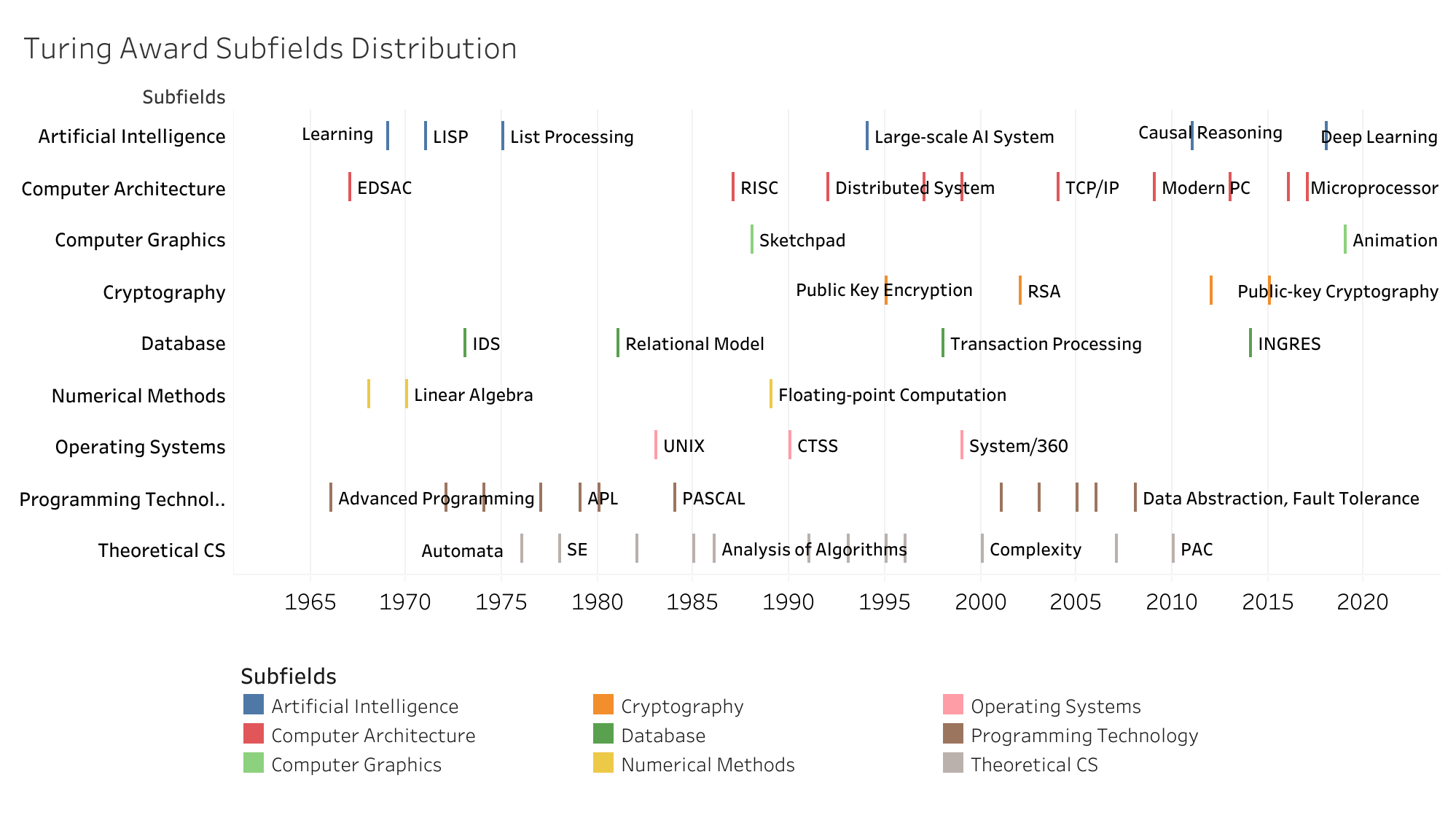}
\caption{The research subjects are classified into nine major sub-fields, each of which is represented in a different row in a different color. All research subjects are arranged in chronological order. A more comprehensive and dynamic presentation made by the Tableau tool can be viewed through the link below\protect\footnotemark. Besides, the complete list can be found in the Appendix.}
\label{fig:4}      
\end{figure}

The Turing Award covers almost all aspects of computer science. There are many ways to classify the sub-research areas of computer science. Here, we refer to the official ACM Turing website of each winner's research subjects. Also, we refer to the classification method of ACMSIG (Special Interest Group), which has 37 main categories. For ease of presentation and analysis, we merged each sub-domain of the Turing Award into several larger domains and merged the sub-domains with a relatively small number of people into approximate domains. Hence, the final nine larger domains are Artificial Intelligence, Computer Architecture, Computer Graphics, Cryptography, Database, Numerical Methods, Operating Systems, Programming Technology, Theoretical CS. The corresponding time distribution of each research subject is shown in Fig. \ref{fig:4}. We can see that Programming Technology and Theoretical CS have respectively won 12 awards, followed by Computer Architecture with 10, Artificial Intelligence with 6, Cryptography and Database with 4, Operating System and Numerical Methods with three, and Computer Graphics with two awards. The full chart of the research field is attached in the Appendix. 

By analyzing the research subject of the Turing Award laureates, we find Turing Award favors original theory and original technology innovation. Many Turing Award winners are "initiators", "founders", and "fathers" of computer theory and technologies. They all show unique ways of thinking and innovative ways, most of which are unprecedented scientific discoveries and principles. For example, in the field of artificial intelligence, the 1969 winner, Marvin Minsky and the 1971 winner John McCarthy are the co-founders of MIT AI Lab, where John McCarthy, recognized as the father of artificial intelligence, was the man who first coined the term "artificial intelligence", opening the field and turning it into a new research area; in the field of automata and computability, the 1976 winners Michael O. Rabin and Dana S. Scott proposed the theory of non-deterministic finite-state automata; the 1993 winners Juris Hartmanis and Richard Edwin Stearns are the main founders of computational complexity theory. In terms of technology, 1967 laureate Maurice V. Wilkes is the builder and designer of the EDSAC, the first computer with an internally stored program; the 1983 winners Ken Thompson and Dennis M. Ritchie are inventors of UNIX; the 1994 winner Edward A. Feigenbaum is known as the father of expert system, together with Raj Reddy, pioneering the design and construction of large scale artificial intelligence systems; the 1997 winner Douglas Engelbart is the inventor of the mouse and other key technologies to help realize the interactive computing; 2004 winners Vinton G. Cerf and Robert E. Kahn designed and implemented the TCP/IP protocol stack.

\footnotetext{\url{https://public.tableau.com/profile/megan.jin\#!/vizhome/researchsubject/Dashboard}}

It can be seen that the Turing Award is a major innovation in the original theory and technology of computers. These major achievements have influenced the development of the computer field. Every major achievement has heralded the advent of a new computer revolution, driving the continuous development of computers to high efficiency, high computability, and practicality.

\subsection{Collaboration Size and Global Collaboration}

\subsubsection{Collaboration Size}
We analyze the collaboration patterns of Turing Award winners from two perspectives, namely, the change in the number of authors in a publication work and the evolution of international collaboration. First of all, we divide the whole period into three sub-periods: 1900-2000, 2001-2010, and 2011-2020. It is because the Turing Award began in 1966, and the laureate's works were finished after 1900. There are 7,965, 2,867, and 2,122 publications, respectively, in these terms. We find from Table 1 that during the period from 1900 to 2000, 38.81\% of the works were independently published, more than half of the works had 2-5 authors, and less than 6\% had more than five collaborators. With the increase of time, the proportion of independent personal works gradually decreased to 16.31\%, the collaborative structure of 2-5 people reached about 63\%, and the proportion of collaborative works with more than five people increased to 20.74\%. This shows that the size of the collaborating group in the computer science field is getting increasing.

\begin{table}[ht]
\caption{Distribution of author structure for the publications by period}
\label{tab:1}       % Give a unique label
\begin{tabular}{lllll}
\hline\noalign{\smallskip}
Period & Single  (ratio)& 2-5 (ratio) & Over 5 (ratio)& Total (ratio) \\
\noalign{\smallskip}\hline\noalign{\smallskip}
1900-2000 & 3,091 (38.81\%) & 4,413 (55.40\%) & 461(5.79\%) & 7,965 (100\%)\\
2001-2010 & 617 (21.52\%) & 1,829 (63.79\%) & 421 (14.68\%) & 2,867 (100\%)\\
2011-2020 & 346 (16.31\%) & 1,336 (62.96\%) & 440 (20.74\%) &  2,122 (100\%)\\
Total  & 4,054 & 7,578 & 1,322 & 12,954 \\
\noalign{\smallskip}\hline
\end{tabular}
\end{table}

\subsubsection{Global Collaboration}
\begin{figure}
	\centering
	\begin{subfigure}{.49\textwidth}
		\includegraphics[width=\textwidth]{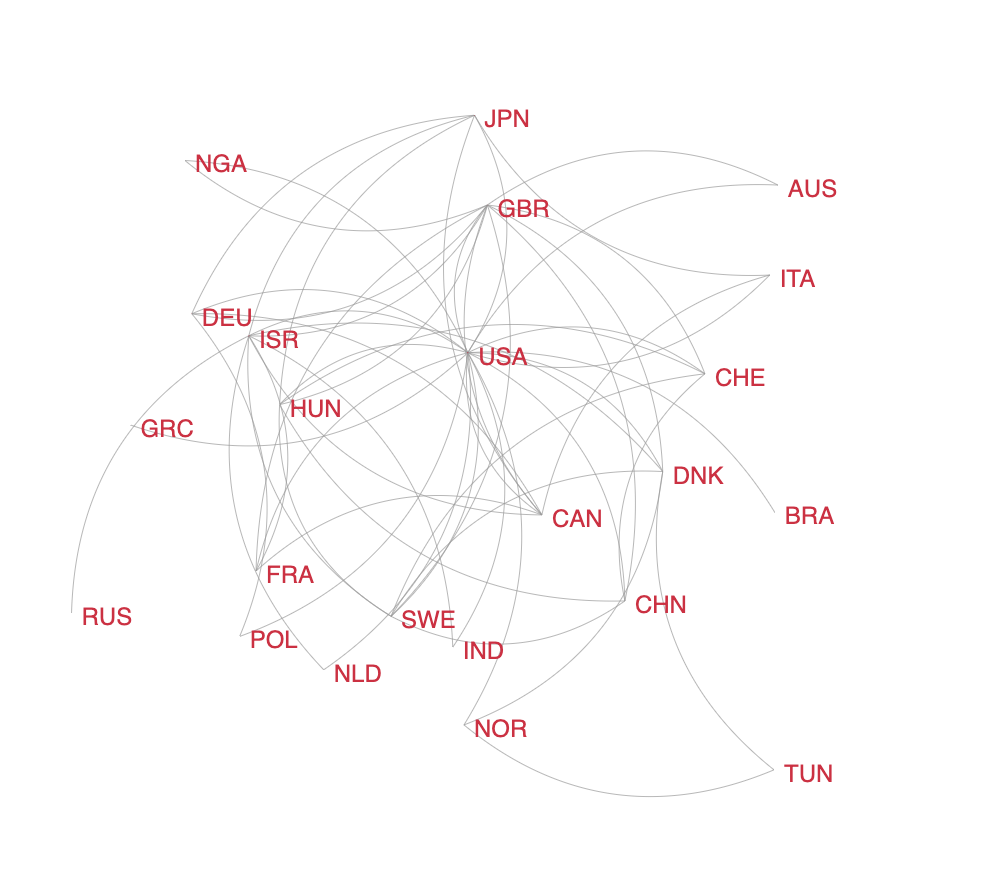}
		\caption{Global collaboration network of the Turing Award laureates from 1900 to 2000.}
	\end{subfigure}
%%%%%%%%%%%%%%
	\begin{subfigure}{.49\textwidth}
		\includegraphics[width=\textwidth]{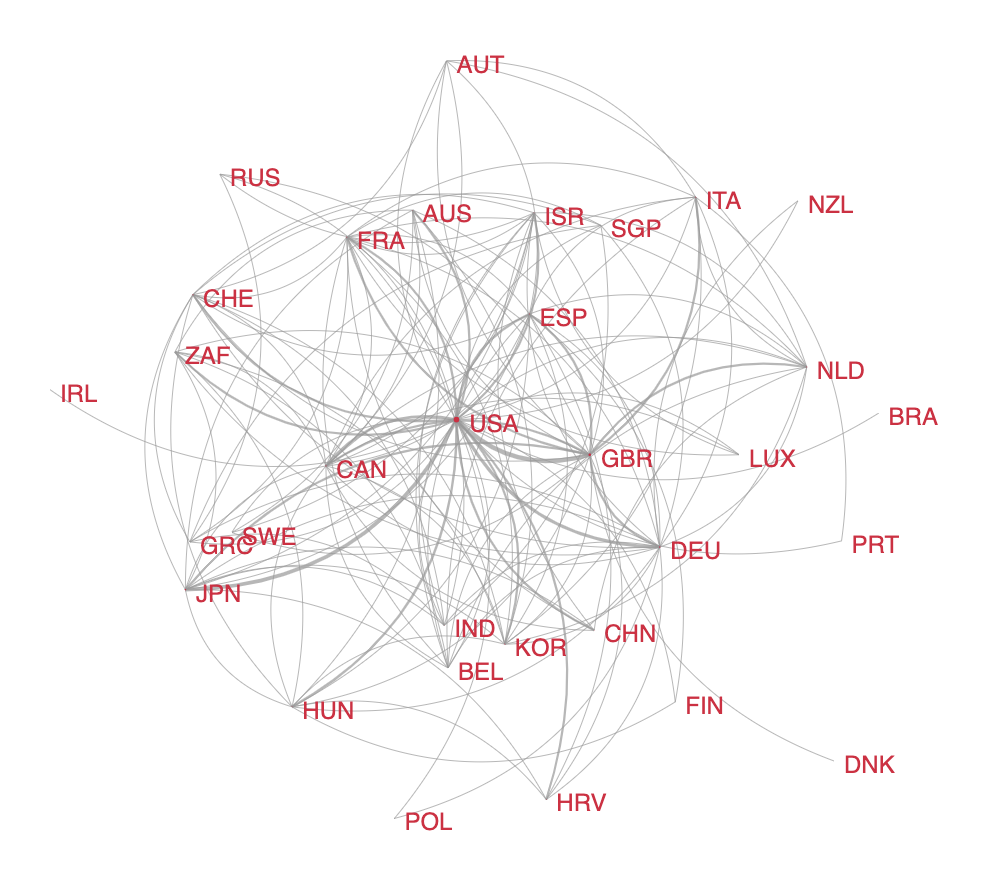}
		\caption{Global collaboration network of the Turing Award laureates from 2001 to 2010.}
	\end{subfigure}
%%%%%%%%%%%%%%
	\begin{subfigure}{.55\textwidth}
		\includegraphics[width=\textwidth]{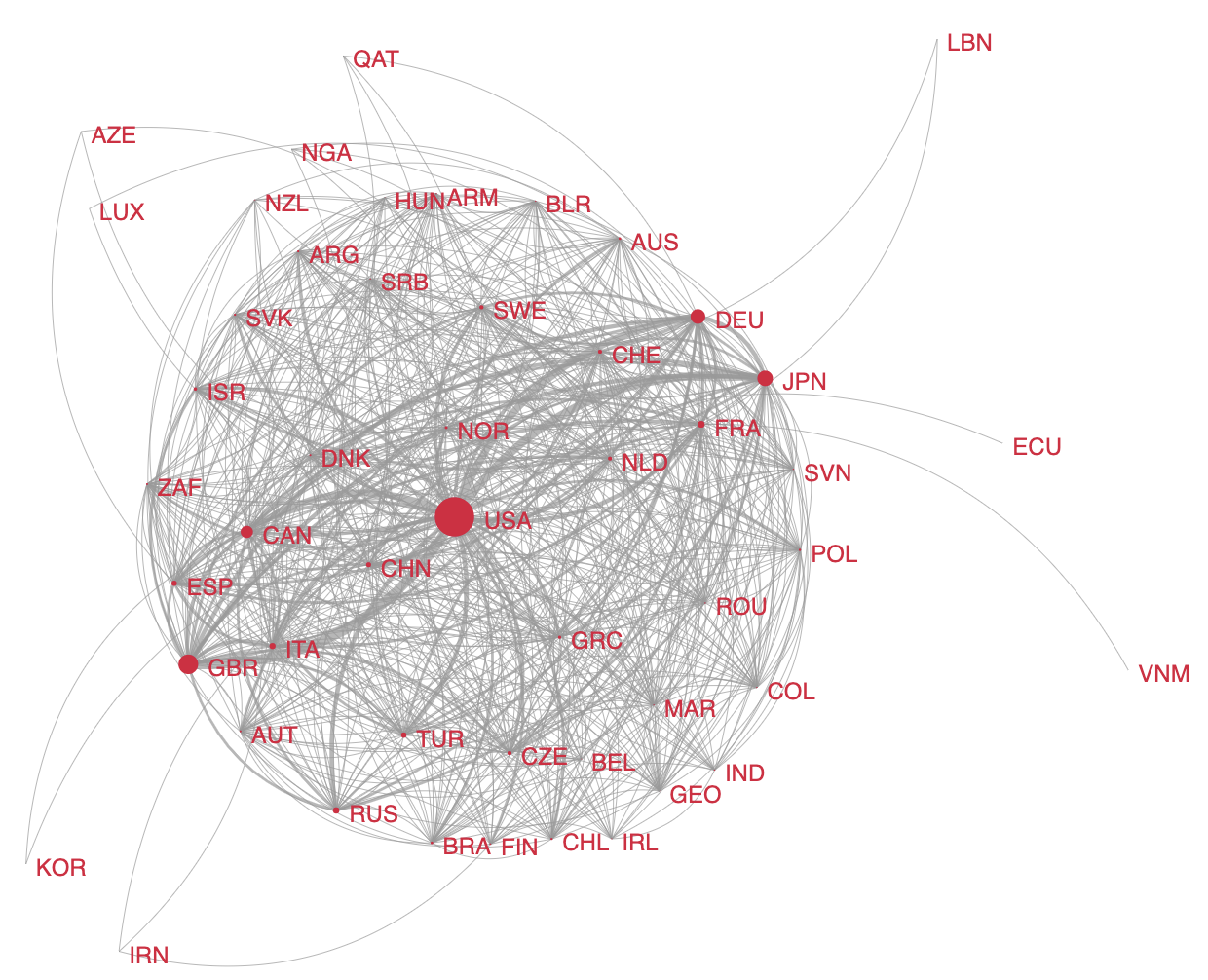}
		\caption{Global collaboration network of the Turing Award laureates from 2011 to 2020.}
	\end{subfigure}
	\caption{The three sub-figures above show the evolution of laureates' global collaboration networks in three different sub-periods. The nodes denote different countries encoded by three-letter national abbreviations, and the thickness of the edges represent the number of collaborations between countries of co-authors' institutional affiliations in the publications of Turing Award winners. The size of the nodes shows the global collaboration numbers of the corresponding country with other countries. The bigger the nodes, the more collaborations with other countries. \scriptsize These figures are generated by ECharts.}
	\label{fig:5}
\end{figure}

To investigate the evolution of international collaboration, in each paper, we extracted a list of the authors' institutional affiliations, then mapped the corresponding countries to each affiliation, and formed the collaboration networks between countries. For example, if a paper is co-authored by five people from four affiliations in the three countries (A, B, C), then there are connections for every two countries, A with B, A with C, and B with C. In this case, A has two global collaborations with B and C, respectively. In Fig. \ref{fig:5}, we can visualize the inter-country collaboration networks in three sub-periods. Here, the nodes denote different countries encoded by three-letter national abbreviations (ISO-3166-1 alpha-3), and the thickness of the edges denote the number of times collaboration has occurred.
These diagrams illustrate the international collaboration was rare before 2000, and then gradually increased from 2001 to 2010, although still very limited. Finally, after 2010, in the recent decade, global collaboration has an explosive increment and has become a new paradigm for academic achievement. 

For quantitative analysis, Table 2 lists the number of publications in each sub-periods, the top 3 countries with the largest number of global collaborations, and the ratio of the number of collaborations to the total number of publications in the corresponding sub-periods. Before 2000, although the number of publications is about three times of the other two sub-periods, global collaboration rarely happened, accounting for only about 5\%. However, it significantly increased from 2001 to 2010. For example, collaboration with the United States was involved in 67.46\% of total publications, followed by the United Kingdom with a quarter and Japan with 15.63\%. Finally, in the last decade, after 2010, there has been an even explosive growth in global collaboration. The number of collaborations with the U.S. accounted for six times the total number of publications, the second-ranked UK three times the total, and Japan 2.4 times the total. Combined with the analysis in Table 1, more and more authors are involved in a publication work, and this kind of collaboration has spread internationally and has become a new paradigm of academic achievement. It is well known that the United States has always been a leader in the international academic arena, and this has been reflected in the Turing Award by winning more than three-quarters of the total. This is also noticeable in Fig. \ref{fig:5} of the global collaboration -- the big red dot, which indicates the U.S. occupied the dominant position in the global collaboration network. Then followed by the UK and Japan, which are also active in the global collaboration network.

\begin{table}
\caption{The number of times that global collaboration has occurred by period}
\label{tab:2}       % Give a unique label
\begin{tabular}{lllll}
\hline\noalign{\smallskip}
Period & N Papers & First (ratio) & Second (ratio)& Third (ratio) \\
\noalign{\smallskip}\hline\noalign{\smallskip}
1900-2000 & 7,965 & USA: 267 (3.35\%) & ISR: 77 (0.97\%) & GBR: 64 (0.80\%) \\
2001-2010 & 2,867 & USA: 1,934 (67.46\%) & GBR: 827 (25.85\%) & JPN: 448 (15.63\%)\\
2011-2020 & 2,122 & USA: 13,046 (614.80\%) & GBR: 6,514 (306.97\%) &  JPN: 5,138 (242.13\%)\\
\noalign{\smallskip}\hline
\end{tabular}
\end{table}

\subsection{Individual Analysis}
We select eight individuals and analyze their publications and corresponding citations in chronological order to look into these elites' research careers. We select four high-productivity researchers, two relatively low-productivity scientists, and two female scientists.

\subsubsection{Productivity VS. Impact}

\begin{figure}[ht]
  \includegraphics[width=1\textwidth]{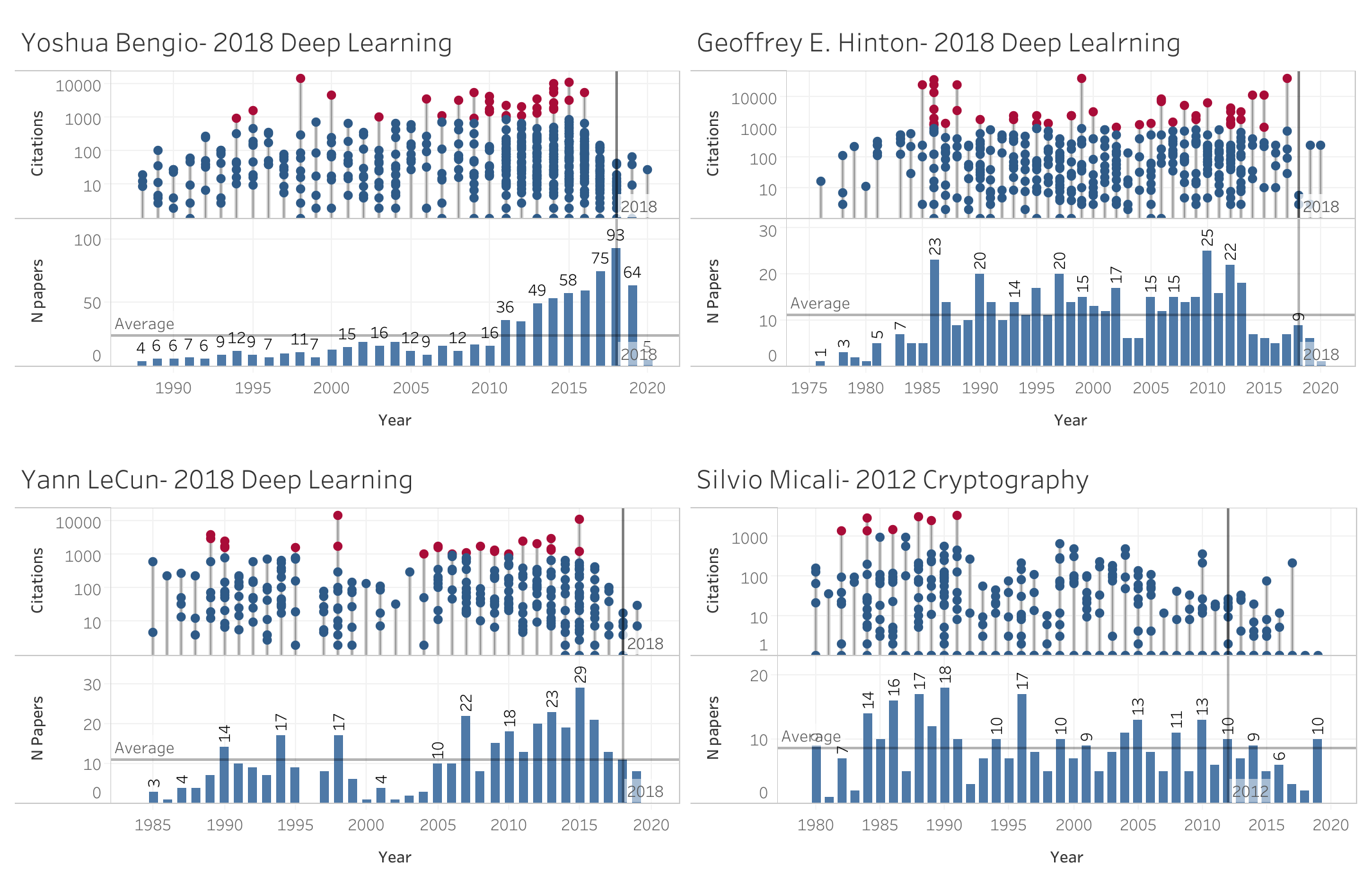}
\caption{These four sub-diagrams show publication details of the four Turing laureates (Bengio (2018), Hinton (2018), LeCun (2018), and Micali (2012)): the number of publications each year (the histogram below each sub-figure) and the number of citations of each paper in the corresponding year (each dot represents the number of citations to each publication). The awarding year and the average number of papers during the researchers' career are also marked in gray lines. The red dots indicate the articles with more than 1,000 citations.}
\label{fig:6}       % Give a unique label
\end{figure}

Fig. \ref{fig:6} shows the outstanding publication records of the three winners in 2018, Yoshua Bengio, Geoffrey Hinton, and Yann LeCun, as well as the winner in 2011 for cryptography, Silvio Micali. These researchers have different productivity patterns. For example, Bengio's pattern has gradually increased from an average of 6 papers per year to a recent average of about 70 per year. At the same time, he is also the most prolific of the 72 Turing Award winners based on the total number of publications. Others either have fluctuating patterns (Hinton and LeCun) or big-headed patterns (Micali). Interestingly, we found that no matter what pattern it looks like, high citation papers -- the red dots, indicating papers with more than 1000 citations are more likely to appear in high-yield years, except that Bengio has started to publish high-impact work continuously from his early research years. In other words, the emergence of high-impact work is positively correlated with high productivity. For example, Hinton's publication volume exceeded the average in 1986-2013, Lecun had higher productivity in 1990, 1994, 1998, and 2005-2016, respectively, and Micali had firm productivity in his early career 1984 to 1991. It can be seen that the red dots appeared in these years. Moreover, this period can occur in early, late, or even a random distribution of one's research career. This is a very interesting discovery, inferring that certain researchers may gain "strong power" within a certain period of time while having considerable productivity and impact of works. This also implies that there are no "bad times" in one's research career.

\subsubsection{Special Case Analysis}

\begin{figure}[ht]
  \includegraphics[width=1\textwidth]{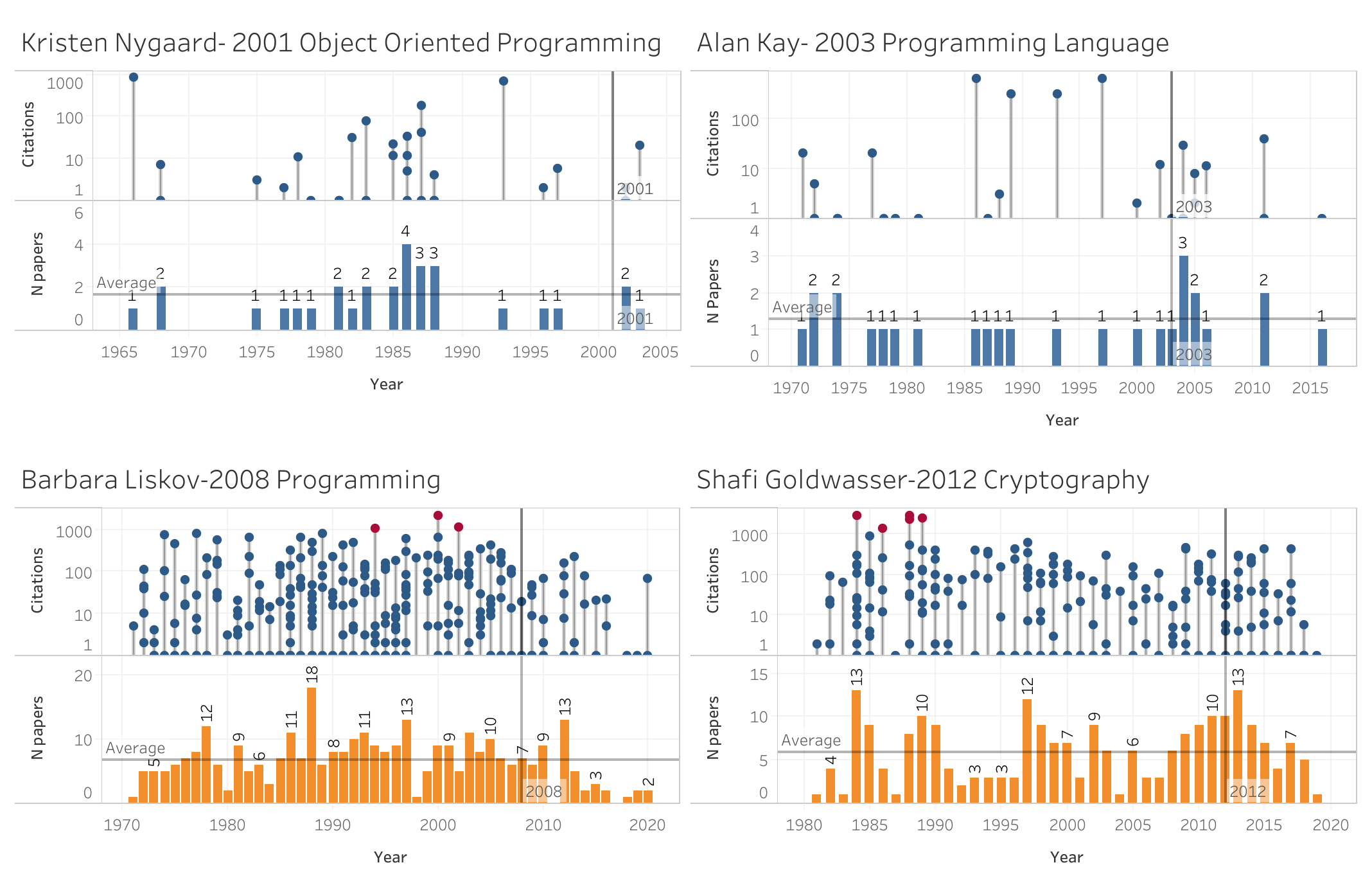}
\caption{These four sub-diagrams show publication details of the four Turing laureates (Nygaard (2001), Kay (2003), Liskov (2008), and Goldwasser's (2012)): the number of publications each year (the histogram below each sub-figure) and the number of citations of each paper in the corresponding year (each dot represents the number of citations to each publication). The awarding year and the average number of papers during the researchers' career are also marked in gray lines. The red dots indicate the articles with more than 1,000 citations.}
\label{fig:7}       % Give a unique label
\end{figure}

Fig. \ref{fig:7} shows some of the relatively low productivity and low citation researchers, Kristen Nygaard, 2001 laureate for ideas to the object-oriented programming and Alan Kay, 2003 laureate for pioneering contemporary Object-oriented programming and also recognized as the father of personal computers. We can see from the figure that in their about 40 years of the research career, the total number of publications did not exceed 30, and the maximum citations of papers were relatively low and never exceeded 1000. If we look at these diagrams alone, their performance may look mediocre, or even worse than ordinary researchers. Nevertheless, they actually spent their lives designing programming languages and developing personal computers. Thus, the evaluation of an individual's contribution is not directly related to the number of publications and their citations. One can make a groundbreaking contribution in a field, but not necessarily in the form of publications.

We also plot two female researchers' trace who won the Turing Award in 2008 and in 2012 in Fig. \ref{fig:7}. From the diagrams, we can see that both of the two female laureates have relatively high productivity with more than 240 publications, respectively, including some high-impact works (red dots). It is worth noting that Barbara Liskov is one of the first women to receive a PhD in computer science in the United States. Her thesis is supervised by John McCarthy, who is another Turing Award winner in 1971, as well as the father of AI. She has contributed remarkably to the theoretical and practical foundations of programming languages and system design, especially related to data abstraction, fault tolerance, and distributed computing. We can also find that compared to the previous two researchers who are also doing research on programming language, her publication production is relatively high. Shafi Goldwasser was a female recipient in 2012. She made significant contributions to encryption, computational complexity, computational number theory, and probabilistic algorithms. Her career includes many landmark papers that have created entire subfields of computer science. In Fig. \ref{fig:7}, she had a persistent and violate production pattern, and the impact also shown in the high-production-high-impact pattern, as we discovered in Fig. \ref{fig:6}.

\section{Related work}

SciSci attempts to uncover the factors that lead to scientific success. For example, most scientists are conservative and tend to focus on their established knowledge \cite{foster2015tradition} since it can remain their relatively high productivity. However, in fact, it is shown that rare innovative combinations of long-distance interdisciplinary knowledge tend to higher citations \cite{lariviere2015long}\cite{kim2016technological}\cite{uzzi2013atypical}. However, most researchers are risk-averse because it is easier to get the reward and recognition, and innovation can be associated with the risk of failure \cite{foster2015tradition}. It appears the same for the funding agencies to evaluate lower scores when they face with new ideas \cite{boudreau2016looking}\cite{yegros2015does} or interdisciplinary proposals \cite{bromham2016interdisciplinary}. Hence, it is necessary to mitigate this conservative trap by urging funding agencies to actively fund risky projects. Moreover, measurement showed that papers written by teams were 6.3 times more likely to get 1000-citations than papers written by individuals \cite{wuchty2007increasing}\cite{lariviere2015team}. It benefited from several aspects. For example, more novel ideas could be integrated \cite{uzzi2013atypical}. Besides, having more collaborators means greater visibility through more co-authors, who may introduce the work to their network \cite{petersen2012persistence}. Another interesting finding is that, on average, the larger the team, the more likely it is to find generous knowledge of the field. In contrast, smaller teams are more likely to discover strange, less common ideas that disrupt the development of the professional field \cite{wu2017large}. Regarding the relationship between creativity and age, the result shows that the most cited papers by scholars can occur at any stage in a person's life \cite{sinatra2016quantifying}\cite{fortunato2018science}, that is, the scientific discovery of a major breakthrough can be any of his or her papers, regardless of age or career stage \cite{sinatra2016quantifying}.

Some scientometric studies have been done on the scientific rewards. For example, a large-scale study covering 3,000 different scientific prizes in diverse disciplines of 10,455 prize-winners worldwide for over 100 years suggests that relatively limited ideas and scholars push the boundaries of science \cite{ma2018scientific}. Therefore, studying these few elites will bring more inspiration to our understanding of scientific evolution. Other studies include research covering all Nobel laureates in three scientific fields up to 2016 \cite{li2020scientific}; research of Nobel Laureates in Physiology or Medicine from 1969 to 2011 \cite{wagner2015nobel}; and the Field Medal, the top award in mathematics, before 1992 was analyzed \cite{borjas2015prizes}. Through the above research on elites in various scientific fields, we can generally conclude that, first, the winners are often energetic producers from the outset, producing works often with higher impact. Moreover, critical works usually occur in their early careers. Secondly, they have more independent works than ordinary researchers. Third, considering the improved reputation and visibility from the prize, we may expect that works produced after the prize gain more impact than those produced before \cite{mazloumian2011citation}, however, counter-intuitively, there is a temporary drop in the productivity and the impact, then it bounces back to their normal previous level. This drop indicates that the scientific community’s attention is not driven by status but the quality of work. Fourth, remarkably, the winners are more likely to change their original research direction after winning. This may be the reason for the temporary drop in productivity and impact. Because changing to a new topic is expensive since it takes extra time to study an unfamiliar field. Finally, in the postprize period, the team collaboration rate has increased, and they are more likely to build bridges across 'small networks'.

We should also note that the above conclusion is not always true in all disciplines. Each discipline has its own unique characteristics and the underlying factors driving its development. If we want to understand the characteristics of a specific discipline, we need to conduct independent research on that particular discipline. Sha Y, et al. \cite{yuan2020science} conducted research focusing on Artificial Intelligence (AI) field. They analyzed the evolution of the subarea of the AI field, AI talents movement, and the change of collaboration patterns. Using data mining technology, they uncover the evolution of AI from the initial focus on mathematical theory to the focus on application in recent years, the flow of AI outstanding talents from developing countries to developed countries, as well as the change of their organizations and the relationship between productivity. Hillebrand \cite{hillebrand2002nobel} concludes that "success is made more likely by an early interest in science, a good education, hard work, mobility (on occasion), as well as a generous portion of luck", after biographically  analyzing Nobel winners in physics.

\section{Conclusion and Discussion}

In this study, we investigated the profiles of 72 Turing Award laureates from 1966 to 2020 and the career prospects of individual researchers. By analyzing Turing Award winners, we can comprehensively understand the evolution of computer science. Our results show that most of the winners have received long-term professional education in the world's top universities, and more than 61\% of the researchers have obtained a degree in mathematics, which shows that mathematics plays a vital role in computer science. The United States is still the main research force in computer science, accounting for about 78\% of the total number of winners. Besides, the recent appearance of female scientists is very eye-catching because there has not been a female laureate since the first 40 years of the Turing Award. Hence, more concrete actions are expected to address the problem of the underrepresentation of women in CS. In general, the productivity and impact of the winners' works are considerable, reaching average publication productivity of about 191.6, with the h-index of 49. While, the deviation is extensive, with coefficients of variation of 88.25\% and 76.19\%, respectively, indicating that these widely-spread metrics are not the primary indicators of the evaluation of the Turing Award. Also, we found an increasing trend in the age of winning awards, and a longer time is needed to receive the award from publishing their first paper. This average age of the award is about 70 years old, and the time to win the award is almost 40 years from their first publications. It has almost reached the limit of the human working age. This may be because, in the modern context, the expansion of fundamental knowledge in a field may increase training requirements, and these scientific breakthroughs need time to verify their value.

In the field of research, programming technology and computer theory received the most awards, accounting for 1/3 of the total awards. Recently, it is more inclined to applications such as artificial intelligence and computer graphics. On the other hand, international collaboration has exploded in the past decade, having more and more collaborators have become a new paradigm for computer science cooperation. In addition, there is no inherent pattern behind their research career. We see many researchers have an outbreak period, that is, the production volume and the impact reach the peak. However, there are also some exceptions. Therefore, predicting the potential of scientists is risky. It is prudent to evaluate and judge the potential of scientific researchers. We can refer to these findings, but we cannot rely entirely on these conclusions because everyone's potential is infinite. Scientific breakthroughs take time. We need to be patient with our researchers, and it is best to provide them with as loose an evaluation environment as possible.

\begin{acknowledgements}
The work is supported by the National Natural Science Foundation of China (NSFC) under Grant No. 61806111, NSFC for Distinguished Young Scholar under Grant No. 61825602 and National Key R\&D Program of China under Grant No. 2020AAA010520002.

\end{acknowledgements}


\bibliographystyle{spmpsci}      % mathematics and physical sciences
\bibliography{scibib}   % name your BibTeX data base

%\begin{thebibliography}{}

%\end{thebibliography}

\begin{table}
\centering
\arrayrulecolor{black}
\begin{tabular}{!{\color{black}\vrule}l!{\color{black}\vrule}c!{\color{black}\vrule}l!{\color{black}\vrule}l!{\color{black}\vrule}} 
\hline
\rowcolor[rgb]{0.651,0.651,0.651} Name & Year & \multicolumn{1}{c!{\color{black}\vrule}}{Subfield}                                   & \multicolumn{1}{c!{\color{black}\vrule}}{Major~Subfield}  \\ 
\hline
EDWIN~E.~CATMULL                       & 2019 & Animation                                                                            & Computer~Graphics                                         \\ 
\hline
PATRICK~M.~HANRAHAN                    & 2019 & Animation                                                                            & Computer~Graphics                                         \\ 
\hline
YOSHUA~BENGIO                          & 2018 & Deep~Learning                                                                        & Artificial~Intelligence                                   \\ 
\hline
GEOFFREY~E~HINTON                      & 2018 & Deep~Learning                                                                        & Artificial~Intelligence                                   \\ 
\hline
YANN~LECUN                             & 2018 & Deep~Learning                                                                        & Artificial~Intelligence                                   \\ 
\hline
JOHN~L~HENNESSY                        & 2017 & Microprocessor                                                                       & Computer~Architecture                                     \\ 
\hline
DAVID~PATTERSON                        & 2017 & Microprocessor                                                                       & Computer~Architecture                                     \\ 
\hline
SIR~TIM~BERNERS-LEE                    & 2016 & World~Wide~Web                                                                       & Computer~Architecture                                     \\ 
\hline
WHITFIELD~DIFFIE                       & 2015 & Public-key~Cryptography                                                              & Cryptography                                              \\ 
\hline
MARTIN~HELLMAN                         & 2015 & Public-key~Cryptography                                                              & Cryptography                                              \\ 
\hline
MICHAEL~STONEBRAKER                    & 2014 & INGRES                                                                               & Database                                                  \\ 
\hline
LESLIE~LAMPORT                         & 2013 & Distributed~and~Concurrent~System                                                    & Computer~Architecture                                     \\ 
\hline
SHAFI~GOLDWASSER                       & 2012 & Complexity-theoretic~foundation                                                      & Cryptography                                              \\ 
\hline
SILVIO~MICALI                          & 2012 & Complexity-theoretic~foundation                                                      & Cryptography                                              \\ 
\hline
JUDEA~PEARL                            & 2011 & Causal~Reasoning                                                                     & Artificial~Intelligence                                   \\ 
\hline
LESLIE~GABRIEL~VALIANT                 & 2010 & PAC                                                                                  & Theoretical~CS                                            \\ 
\hline
CHARLES~P.~(CHUCK)~THACKER             & 2009 & Modern~PC                                                                            & Computer~Architecture                                     \\ 
\hline
BARBARA~LISKOV                         & 2008 & Data~Abstraction,~Fault~Tolerance                                                    & Programming~Technology                                    \\ 
\hline
EDMUND~MELSON~CLARKE                   & 2007 & Model~Checking                                                                       & Theoretical~CS                                            \\ 
\hline
E.~ALLEN~EMERSON                       & 2007 & Model~Checking                                                                       & Theoretical~CS                                            \\ 
\hline
JOSEPH~SIFAKIS                         & 2007 & Model~Checking                                                                       & Theoretical~CS                                            \\ 
\hline
FRANCES~("FRAN")~ELIZABETH~ALLEN       & 2006 & Compilers                                                                            & Programming~Technology                                    \\ 
\hline
PETER~NAUR                             & 2005 & ALGOL~60                                                                             & Programming~Technology                                    \\ 
\hline
VINTON~(“VINT”)~GRAY~CERF              & 2004 & TCP/IP                                                                               & Computer~Architecture                                     \\ 
\hline
ROBERT~(“BOB”)~ELLIOT~KAHN             & 2004 & TCP/IP                                                                               & Computer~Architecture                                     \\ 
\hline
ALAN~KAY                               & 2003 & Object~Oriented~Programming                                                          & Programming~Technology                                    \\ 
\hline
LEONARD~(LEN)~MAX~ADLEMAN              & 2002 & RSA                                                                                  & Cryptography                                              \\ 
\hline
RONALD~(RON)~LINN~RIVEST               & 2002 & RSA                                                                                  & Cryptography                                              \\ 
\hline
ADI~SHAMIR                             & 2002 & RSA                                                                                  & Cryptography                                              \\ 
\hline
OLE-JOHAN~DAHL                         & 2001 & Object~Oriented~Programming                                                          & Programming~Technology                                    \\ 
\hline
KRISTEN~NYGAARD                        & 2001 & Object~Oriented~Programming                                                          & Programming~Technology                                    \\ 
\hline
ANDREW~CHI-CHIH~YAO                    & 2000 & Complexity                                                                           & Theoretical~CS                                            \\ 
\hline
FREDERICK~("FRED")~BROOKS              & 1999 & System/360                                                                           & Computer~Architecture                                     \\ 
\hline
FREDERICK~("FRED")~BROOKS              & 1999 & System/360~                                                                          & Operating~Systems                                         \\ 
\hline
JAMES~("JIM")~NICHOLAS~GRAY            & 1998 & Transaction~Processing                                                               & Database                                                  \\ 
\hline
DOUGLAS~ENGELBART                      & 1997 & Interactive~Computing                                                                & Computer~Architecture                                     \\ 
\hline
AMIR~PNUELI                            & 1996 & Temporal~Logic                                                                       & Theoretical~CS                                            \\ 
\hline
MANUEL~BLUM                            & 1995 & Public~Key~Encryption                                                                & Cryptography                                              \\ 
\hline
MANUEL~BLUM                            & 1995 & Computational~Complexity~                                                            & Theoretical~CS                                            \\ 
\hline
EDWARD~A~("ED")~FEIGENBAUM             & 1994 & Large-scale~AI~System                                                                & Artificial~Intelligence                                   \\ 
\hline
DABBALA~RAJAGOPAL~("RAJ")~REDDY        & 1994 & Large-scale~AI~System                                                                & Artificial~Intelligence                                   \\ 
\hline
JURIS~HARTMANIS                        & 1993 & Computational~complexity                                                             & Theoretical~CS                                            \\ 
\hline
RICHARD~("DICK")~EDWIN~STEARNS         & 1993 & Computational~complexity                                                             & Theoretical~CS                                            \\ 
\hline
BUTLER~W~LAMPSON                       & 1992 & Distributed~System                                                                   & Computer~Architecture                                     \\ 
\hline
ARTHUR~JOHN~ROBIN~GORELL~MILNER        & 1991 & LCF,~ML                                                                              & Theoretical~CS                                            \\ 
\hline
FERNANDO~J~("CORBY")~CORBATO           & 1990 & CTSS                                                                                 & Operating~Systems                                         \\ 
\hline
WILLIAM~(“VELVEL”)~MORTON~KAHAN        & 1989 & Floating-point~Computation                                                           & Numerical~Methods                                         \\ 
\hline
IVAN~SUTHERLAND                        & 1988 & Sketchpad                                                                            & Computer~Graphics                                         \\ 
\hline
JOHN~COCKE                             & 1987 & RISC                                                                                 & Computer~Architecture                                     \\ 
\hline
JOHN~E~HOPCROFT                        & 1986 & Analysis~of~Algorithms                                                               & Theoretical~CS                                            \\ 
\hline
ROBERT~(BOB)~ENDRE~TARJAN              & 1986 & Analysis~of~Algorithms                                                               & Theoretical~CS                                            \\ 
\hline
RICHARD~("DICK")~MANNING~KARP          & 1985 & Combinatorial~Algorithms                                                             & Theoretical~CS                                            \\ 
\hline
NIKLAUS~E.~WIRTH                       & 1984 & PASCAL                                                                               & Programming~Technology                                    \\ 
\hline
DENNIS~M.~RITCHIE                      & 1983 & UNIX                                                                                 & Operating~Systems                                         \\ 
\hline
KENNETH~LANE~THOMPSON                  & 1983 & UNIX                                                                                 & Operating~Systems                                         \\ 
\hline
STEPHEN~ARTHUR~COOK                    & 1982 & Computational~Complexity                                                             & Theoretical~CS                                            \\ 
\hline
EDGAR~F.~("TED")~CODD                  & 1981 & Relational~Model~                                                                    & Database                                                  \\ 
\hline
C.~ANTONY~R.~HOARE~                    & 1980 & \begin{tabular}[c]{@{}l@{}}Programming~Language~Definition~\\and~Design\end{tabular} & Programming~Technology                                    \\ 
\hline
KENNETH~E.~("KEN")~IVERSON             & 1979 & APL                                                                                  & Programming~Technology                                    \\ 
\hline
ROBERT~(BOB)~W~FLOYD                   & 1978 & Software Engineering                                                                                   & Theoretical~CS                                            \\ 
\hline
JOHN~BACKUS                            & 1977 & High~Level~Programing~System                                                         & Programming~Technology                                    \\ 
\hline
MICHAEL~O.~RABIN                       & 1976 & Automata                                                                             & Theoretical~CS                                            \\ 
\hline
DANA~STEWART~SCOTT                     & 1976 & Automata                                                                             & Theoretical~CS                                            \\ 
\hline
ALLEN~NEWELL                           & 1975 & List~Processing                                                                      & Artificial~Intelligence                                   \\ 
\hline
HERBERT~ALEXANDER~SIMON                & 1975 & List~Processing                                                                      & Artificial~Intelligence                                   \\ 
\hline
DONALD~("DON")~ERVIN~KNUTH             & 1974 & Programming~Language~Design                                                          & Programming~Technology                                    \\ 
\hline
CHARLES~WILLIAM~BACHMAN                & 1973 & IDS                                                                                  & Database                                                  \\ 
\hline
EDSGER~WYBE~DIJKSTRA                   & 1972 & High~Level~Programing~Language                                                       & Programming~Technology                                    \\ 
\hline
JOHN~MCCARTHY                          & 1971 & LISP                                                                                 & Artificial~Intelligence                                   \\ 
\hline
JAMES~HARDY~("JIM")~WILKINSON          & 1970 & Linear~Algebra                                                                       & Numerical~Methods                                         \\ 
\hline
MARVIN~MINSKY~                         & 1969 & Learning                                                                             & Artificial~Intelligence                                   \\ 
\hline
RICHARD~W.~HAMMING                     & 1968 & Automatic~Coding~System                                                              & Numerical~Methods                                         \\ 
\hline
MAURICE~V.~WILKES~                     & 1967 & EDSAC                                                                                & Computer~Architecture                                     \\ 
\hline
ALAN~J~PERLIS                          & 1966 & Advanced~Programming                                                                 & Programming~Technology                                    \\
\hline
\end{tabular}
\arrayrulecolor{black}
\end{table}

\end{document}